\documentclass[%
 aip,
 amsmath,amssymb,
 reprint,%
]{revtex4-1}

\usepackage{graphicx}
\usepackage{dcolumn}
\usepackage{bm}

\usepackage[utf8]{inputenc}
\usepackage[T1]{fontenc}
\usepackage{mathptmx}

\begin{document}

\preprint{AIP/123-QED}

\title{Progressive-Growing of Generative Adversarial Networks for Metasurface Optimization}

\author{Fufang Wen}
\altaffiliation{These two authors contribute equally to this work.}
\affiliation{Department of Electrical Engineering, Columbia University, NY, 10027, USA}

\author{Jiaqi Jiang}
\altaffiliation{These two authors contribute equally to this work.}
\affiliation{Department of Electrical Engineering, Stanford University, CA, 94305, USA}

\author{Jonathan A. Fan}
\email{jonfan@stanford.edu}
\affiliation{Department of Electrical Engineering, Stanford University, CA, 94305, USA}

\date{\today}

\begin{abstract}

Generative adversarial networks, which can generate metasurfaces based on a training set of high performance device layouts, have the potential to significantly reduce the computational cost of the metasurface design process.  However, basic GAN architectures are unable to fully capture the detailed features of topologically complex metasurfaces, and generated devices therefore require additional computationally-expensive design refinement.  In this Letter, we show that GANs can better learn spatially fine features from high-resolution training data by progressively growing its network architecture and training set.  Our results indicate that with this training methodology, the best generated devices have performances that compare well with the best devices produced by gradient-based topology optimization, thereby eliminating the need for additional design refinement.  We envision that this network training method can generalize to other physical systems where device performance is strongly correlated with fine geometric structuring.

\end{abstract}

\maketitle

\section{Introduction}
Metasurfaces are nanostructured electromagnetic media with responses tailored by structure geometry \cite{Yu14,Kildishev13,Kuznetsov16}. They are an emergent technology and are relevant to a wide scope of applications, including those in imaging \cite{Colburn14}, sensing \cite{Tittl18}, polarization control \cite{Arbabi15}, and holography \cite{Zheng15}.  The identification of an effective and computationally efficient design method for high performance metasurfaces remains an open-ended problem \cite{Campbell19}. Optimization-based methods, ranging from gradient-based topology optimization \cite{Molesky18,Jensen11} to genetic algorithms \cite{Jafar-Zanjani18}, have become promising approaches to designing high-performance nanophotonic devices such as metagratings \cite{Sell18,Yang18,Jiang19(1),Jiang19(2)}, metasurfaces \cite{Sell17,Lin18}, and in-plane nanophotonic devices \cite{Xiao16,Hughes18,Piggott15,Lalau-Keraly13}. However, these methods are computationally expensive, making their scaling to large ensembles of topologically-complex devices very costly and in some limits intractable.

Various concepts based on machine learning have been proposed to mitigate the computational bottleneck of conventional optimization approaches \cite{Liu18,Inampudi18}. One promising concept is the generative adversarial network (GAN) \cite{Goodfellow14, Jiang19_acsnano, wenshan2018, junsuk2019}, which learns from images of high-performance metasurfaces and can generate high resolution device layouts with topologically-complex features.  While this approach requires the creation of a computationally expensive training set, this expense is a one time cost, and a trained GAN can generate a diversity of device layouts within seconds.  In a recent study \cite{Jiang19_acsnano}, we trained a GAN from images of gradient-based topology optimized metagratings, which are periodic metasurfaces that selectively diffract incident light to a desired diffraction order.  Our GAN possessed a conventional deep network architecture, and it was successful in generating ensembles of topologically-complex metagratings operating across a range of wavelengths and deflection angles.  However, the performances of the best GAN-generated devices were consistently lower than that those in the training set, and additional iterations of computationally costly gradient-based optimization were required to improve these devices.

There are at least two reasons why conventional GAN generators are not able to learn the detailed spatial features of the training set.  One is because the design space of the training set is vast and the GAN generator is unable to learn detailed features during training \cite{Odena17}. Another is that the use of a small training set makes it easy for the discriminator to overfit.  To address these problems, researchers in the computer vision community have proposed the progressive growth of GANs (PGGANs), which is a new GAN training scheme that supports improved training stability and the ability to capture spatially fine features from a high-resolution training set \cite{Karras18}.  In this scheme, the PGGAN architecture operates with low spatial resolution during initial training iterations and focuses on learning spatially coarse features from the training set.  Additional neural layers are then progressively added, at which point the PGGAN focuses on learning finer-scale details from the training set.  Given the success of these techniques to improve the generation of high-resolution images, such as the faces of people, these networks serve as plausible candidate solutions to improving the generation of physical devices such as metasurfaces.

\begin{figure*}[htp]
\centering
\includegraphics[width=\linewidth]
{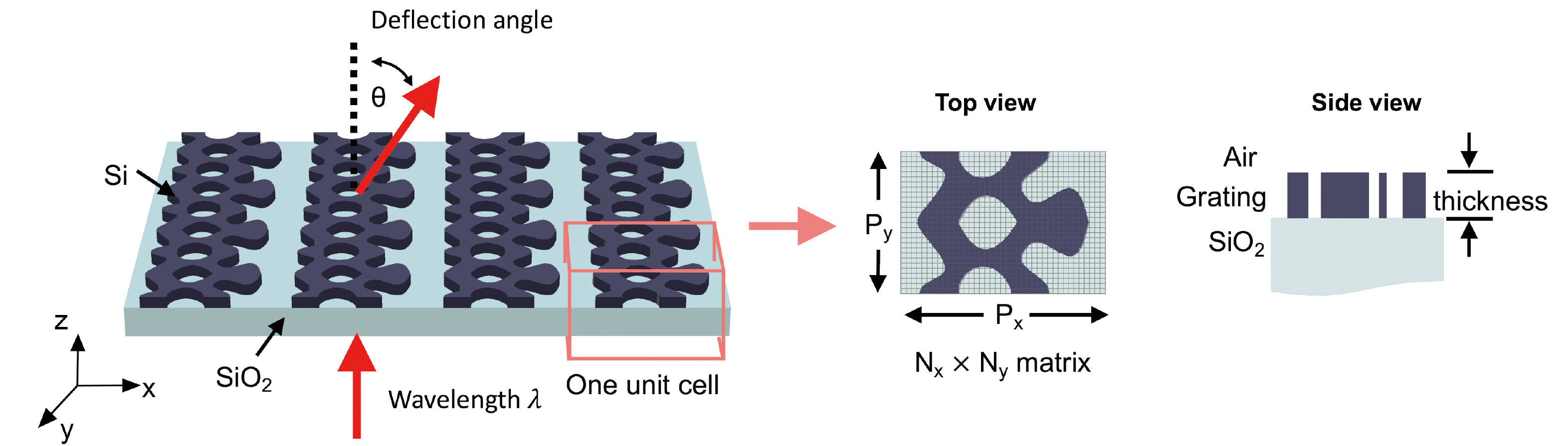}
\caption{The 3D metagratings consist of freeform silicon patterns and deflect normally-incident light to the +1 diffraction order. For each period, the metagrating is described by an $N_{x}$ by $N_{y}$ matrix, and each pixel possesses a refractive index value of either silicon or air. The width of the period along the $x$ direction is $\lambda/\sin(\theta)$ and the width of the period along the $y$ direction is set to be $\lambda/2$. Reflection symmetry is enforced in the $y$ direction. 
}
\label{fig1}
\end{figure*}

In this Letter, we show that the performance of the GAN generator for metasurfaces can be dramatically enhanced through the progressive growth of the GAN architecture and training set during training.  In addition to increasing the resolution of the GAN throughout the network training process, we progressively augment the training set by identifying high-performance GAN-generated metasurfaces using an electromagnetic simulator and adding them to the training set, while removing relatively lower performance devices.  As a model system, we apply our progressive growth scheme to train a conditional metagrating PGGAN that can produce ensembles of devices operating across a range of wavelengths and deflection angles.  While our network training scheme incurs a large one-time computational cost, the final PGGAN is capable of generating devices with efficiencies comparable to and sometimes better than the best devices in the training set, which eliminates the need for additional computationally-expensive device refinement.



\section{Problem setup and overview of network training}
We focus on designing topologically complex metagratings that deflect normally incident TE-polarized light to the +1 diffraction order for a range of outgoing angles (35 degrees to 85 degrees) and operating wavelengths (500 nm to 1100 nm).  A schematic of the device is summarized in Fig. \ref{fig1}. The metagratings comprise 325 nm-thick polycrystalline silicon on an SiO$_2$ substrate, and devices are represented as images of single grating periods with dimensions of $64 \times 128$ pixels. Each pixel in the image has a value of either 0 or 1, which represents the refractive index of air or polycrystalline silicon, respectively.  Mirror symmetry along the $y$ direction is enforced to simplify the design space. The deflection efficiency is defined as the intensity of light deflected to the desired angle $\theta$, normalized to the incident light intensity in glass. 
Prior to PGGAN training, we create a training set containing 400 high-efficiency metasurfaces, each produced by performing 350 iterations of gradient-based topology optimization on an initially random dielectric distribution. These devices sparsely sample the design parameter space: the training set devices sample the wavelength space in increments of 200 nm and the deflection angle space in increments of 10 degrees.  To evaluate the performance of the training set devices and those generated by the PGGAN, we use the Rigorous Coupled-Wave Analysis (RCWA) electromagnetic simulator \cite{Hugonin05}.

Our PGGAN comprises two neural networks, a generator and discriminator.  The inputs to the generator include the operating wavelength $\lambda$, deflection angle $\theta$, and an 8-dimensional uniformly distributed random noise vector $z$, and its output is an image of the device.  Given a distribution of noise values as inputs, the outputs are a diverse distribution of devices.  The discriminator is a classifier that attempts to distinguish whether a presented input image is from the generator or the training set.  Architecturally, the generator contains fully connected layers followed by deconvolution layers, while the discriminator contains convolution layers followed by fully connected layers.

The training process can be described as a competition between the generator and discriminator.  The discriminator aims to successfully distinguish between generated and training set devices, while the generator aims to fool the discriminator by generating devices mimicking the training set. The generator and discriminator train through an iterative process in alternating steps, and upon training completion, the final generator will have learned the underlying topological features from high-efficiency devices in the training set and be able to generate layouts with high-efficiency features.  During training, both the generator and discriminator employ the Adam optimizer with a batch size of 256, learning rate of 0.001, $\beta_1$ of 0, and $\beta_2$ of 0.9. We employ Wasserstein loss with a gradient penalty with lambda of 10 for the discriminator loss \cite{Arjovsky17}.  

\begin{figure*}[htp]
\centering
\includegraphics[width=0.95\linewidth, height=\linewidth]{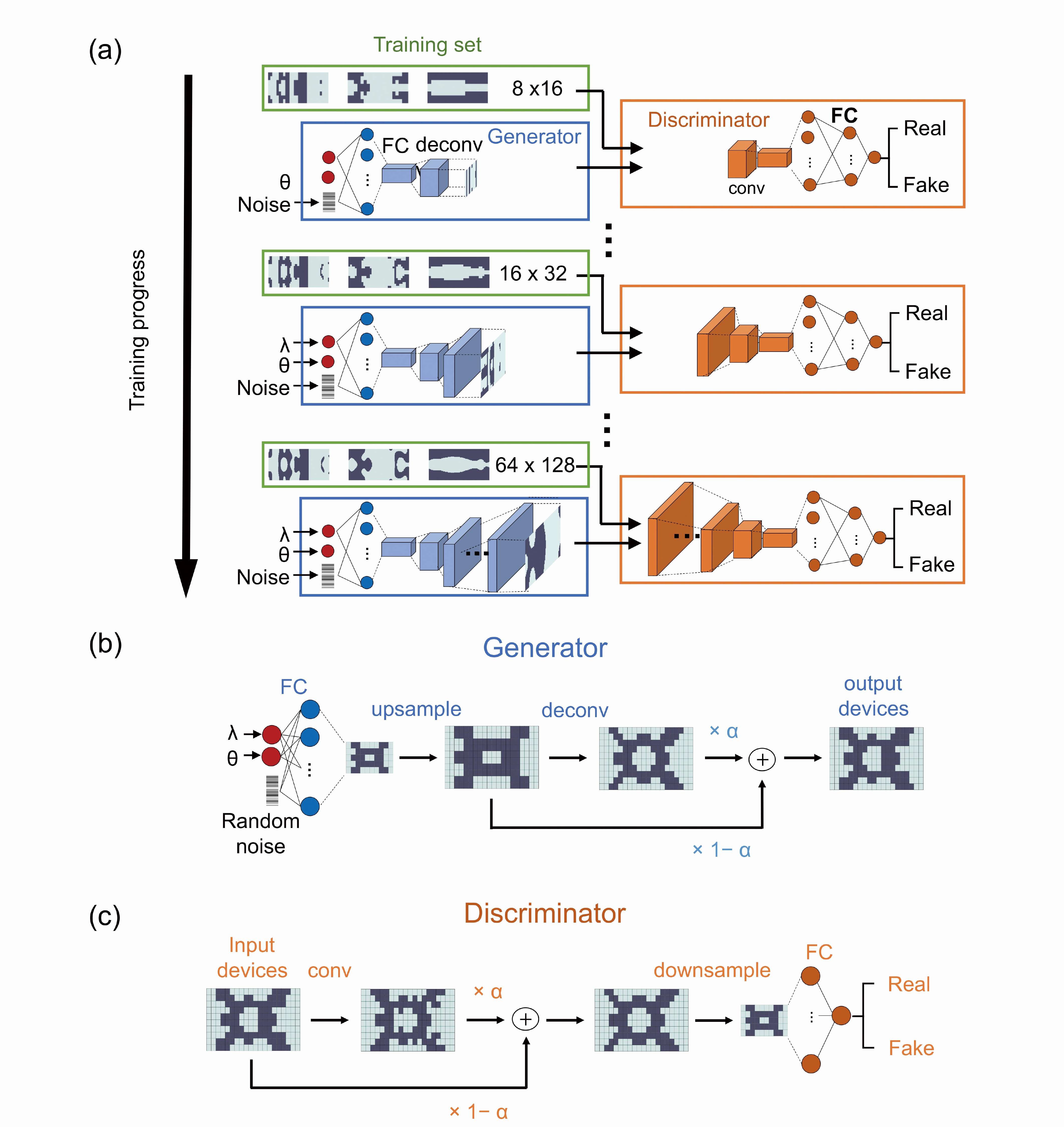}
\caption{(a) Schematics of the network training process, in which the generative model, discriminative model, and network resolution progressively grow over the course of training.  The training starts by processing training set devices with a resolution of $8 \times 16$ pixels, downsampled from full resolution devices of $64 \times 128$ pixels, and the device and network resolution gradually increase to the full resolution of the original training set resolution.  (b) Generator network transitioning from a lower to higher spatial resolution. (c) Discriminator network transitioning from a lower to higher spatial resolution.  During the resolution transition, the impact of new layers is progressively increased in the network architecture by linearly tuning the weight $\alpha$ from 0 to 1.}
\label{fig2}
\end{figure*}

\subsection{Progressive growing of the network architecture}
To improve the stability of the training process and enhance the capabilities of both the generator and discriminator, we progressively grow the resolution of the PGGAN over the course of training \cite{Karras18}.  Schematics of the PGGAN architecture at different stages of network training are presented in Fig. \ref{fig2}(a).  The initial network architecture processes devices with a resolution of only $8 \times 16$ pixels and trains from downsampled images of the training set.  Our use of low-resolution devices and networks has three main implications.  First, it allows the network to focus on learning the large-scale topological features from the training set.  Second, it restricts the design space to a much lower dimension, which improves the overall accuracy of the network training process itself.  Third, training at lower spatial resolutions leads to dramatic reductions in computational cost.

Once the network has undergone sufficient training with these low-resolution devices, the network and device resolution increase to $16 \times 32$ pixels, and then later to $32 \times 64$ pixels and finally to $64 \times 128$ pixels.  During each increase in resolution, an additional convolution and deconvolution layer is added to the generator and discriminator, respectively, each with dimensions matching the new device resolution.  With these added layers, the PGGAN accurately learns and captures higher spatial resolution features in the training set.

\begin{figure*}[ht]
\centering
\includegraphics[width=\linewidth]
{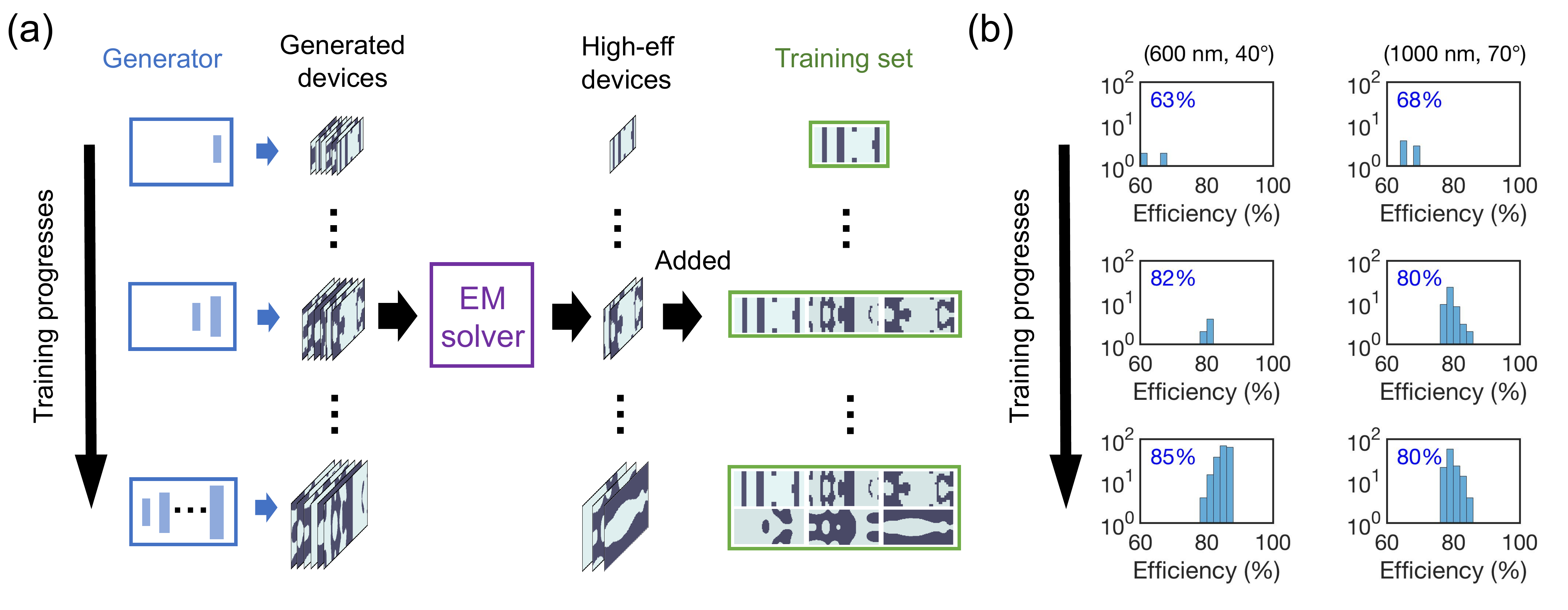}
\caption{(a) Schematic of the growth of the training set over the course of network training.  High efficiency devices produced from the generator are identified and added to the training set throughout the training process. (b) The evolution of efficiency distributions from the training set over the course of network training for two representative wavelength-deflection angle pairs: (600 nm, 40$^\circ$) and (1000 nm, 70$^\circ$).  These efficiency distributions are sampled from three network training moments: the first training set update, the sixth training set update, and the final training set update. The average efficiency of the training set for each wavelength-deflection angle pair is denoted in each plot.
}
\label{fig3}
\end{figure*}

During the transition moments in the training process when the spatial resolution of the network increases, care must be taken to ensure that this increase in spatial resolution does not destabilize the network.  These instabilities can arise because the convolution and deconvolution layers that are added to the PGGAN during these moments are untrained and possess random weights.  To address this issue, the impact of the added convolution and deconvolution layers are gradually incorporated into the PGGAN in a manner illustrated in Fig. \ref{fig2}b.  Consider the PGGAN generator.  When a new deconvolution layer is added to the network, the device at the output of this layer $X_{new}$ is the weighed superposition of two images.  The first is $X_{deconv}$, which is the device outputted from the prior deconvolution layer, upsampled to match the spatial resolution of the new deconvolution layer, and then processed by the deconvolution layer.  The second is the device $X_{upsampled}$, which is the upsampled device.  $X_{new}$ relates to $X_{upsampled}$ and $X_{deconv}$ with the following expression:
\begin{equation}
X_{new} = \alpha*X_{deconv}+(1-\alpha)*X_{upsampled}
\label{eq:refname1}
\end{equation}
Over the course of 5000 iterations of network training, $\alpha$ linearly increases from zero to one.  Initially, $\alpha$ is zero and the untrained deconvolution layer does not contribute to the generated device pattern.  As training progresses and $\alpha$ increases, this layer begins to properly learn and capture spatially fine device features and contributes more to $X_{new}$.  After 5000 iterations, the generated device patterns are exclusively generated from the new deconvolution layer.  5000 additional training iterations are performed after $\alpha$ is set to one to further stabilize the network, after which the network resolution is increased again and the process is repeated.  The PGGAN discriminator evolves in a similar way, except that convolutional layers are progressively added to the network instead of deconvolutional layers.  The downsampling operation in the discriminator is performed using average pooling.

\subsection{Progressive growing of the training set}
In addition to progressively growing the network architecture, we progressively grow the training set by adding relatively high-performance PGGAN-generated data to the training set while removing relatively low-performance training data throughout the training process.  The training set growth process is illustrated in Fig. \ref{fig3}a.  Methods to improve the training set are essential to train the best generator possible, as a properly trained generator produces devices with topological features mimicking the training set.  As such, the only way for the generator to consistently generate devices with greater geometric diversity and enhanced performance compared to the initial training set is to evolve and augment the training set.

The generation of high-performance GAN-based devices is stochastic and due to the use of random noise as an input to the generator, which enables a diverse distribution of designs in the high-dimensional design space to be produced.  To identify high-performance devices suitable for training set augmentation, we simulate ensembles of generated devices using the RCWA EM solver and filter out the high-efficiency devices. For each training set growth step, we add high-efficiency devices to the training set while deleting the low-efficiency one (see Appendix A).
\begin{figure*}[ht]
\centering
\includegraphics[width=\linewidth]
{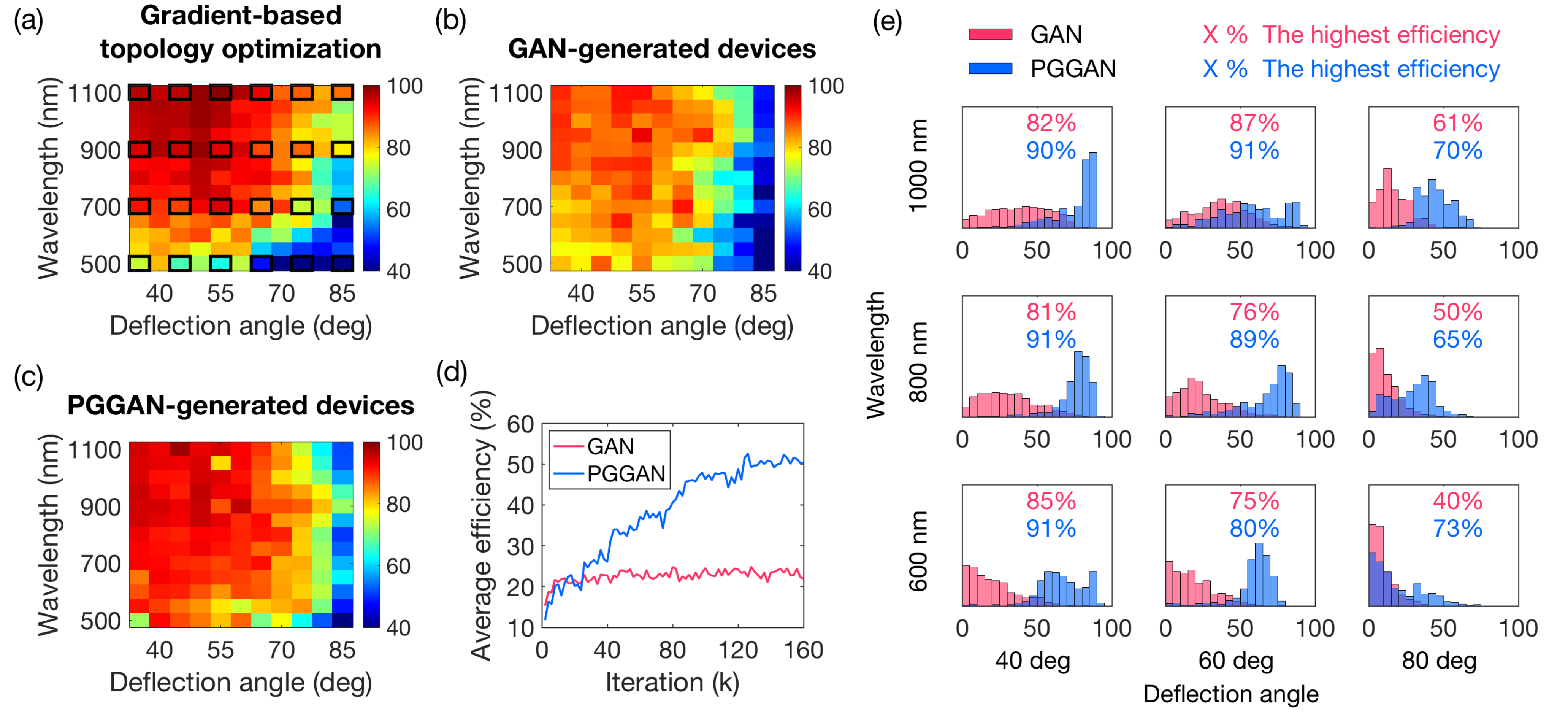}
\caption{Summary of PGGAN performance.  (a-c) Plot of the highest efficiency devices at each wavelength-deflection angle pair for devices designed using (a) gradient-based topology optimization, (b) GAN, and (c) PGGAN.  The initial training set used for the GAN and PGGAN have parameters denoted by the solid black boxes in (a).  (d) Average generated device efficiency as a function of training iteration for GAN (red) and PGGAN (blue). (e) Representative efficiency histograms of GAN-generated devices (red) and PGGAN-generated devices (blue). 500 devices are generated from the GAN and PGGAN for each wavelength-deflection angle pair. The highest efficiencies in each histogram are denoted.}
\label{fig4}
\end{figure*}

Histograms of training set device efficiencies for 2 representative wavelength-deflection angle pairs, over the course of network training, are summarized in Fig. \ref{fig3}b. As training progresses, more high performance devices are added to the training set and both the quantity and average efficiency of the devices in the training set increases.  The average efficiency of the training set converges to an asymptotic value after fifteen training set updates, or 160,000 training iterations, which sets the total number of PGGAN training iterations.

Our ability to improve the PGGAN using PGGAN-generated devices leads to a positive feedback loop between network and training set augmentation: improvements to the training set leads to enhanced discriminator and generator performance, while improvements to the generator lead to enhanced generation of high efficiency metasurfaces that are added to the training set. By alternating these two processes, the network not only learns the desired features from the training set, it also explores and interpolates better topological features in the design space.

\section{Results \& discussion}

We produce final device designs from our PGGAN by generating 2000 devices for a given wavelength-deflection angle pair, evaluating the efficiencies of these devices using the RCWA EM solver, and taking the highest efficiency device.  The operating device parameter space includes wavelengths ranging from 500 nm to 1100 nm, in increments of 50 nm, and deflection angles ranging from 35 degrees to 85 degrees, in increments of 5 degrees.  As benchmarks, we also design devices in the following two ways: we design a total of 8000 devices using gradient-based topology optimization \cite{sell2017large} and select the best device for a given wavelength-deflection angle pair, and we train a basic GAN without progressive growth and filter for high-performance generated devices in the same manner above.  Plots that summarize the efficiencies of devices designed using these three methods are shown in Figs. 4(a)-4(c).

A comparison between Figs. \ref{fig4}(a) and \ref{fig4}(c) shows that the best devices generated by the PGGAN have efficiencies similar to or even better than those produced from gradient-based topology optimization.  Statistically, for $50\%$ of the wavelength-deflection angle pairs, the best PGGAN-generated devices outperform those from gradient-based topology optimization.  Furthermore, when averaging all of the efficiency values from the plots in Fig. \ref{fig4}(a) and \ref{fig4}(c), the average efficiency value from the PGGAN is $1.4 \%$ higher than that from gradient-based topology optimization. The high efficiencies for devices generated at all wavelength-deflection angle pairs, including those not covered in the original training set, indicates that our PGGAN strategy can properly generalize device design across the full wavelength and deflection angle parameter space without significant overfitting.

A comparison between Figs. \ref{fig4}(b) and \ref{fig4}(c) shows that the PGGAN outperforms the basic GAN by a wide margin, further demonstrating the effectiveness of the progressive growth strategy.  For $89\%$ of the wavelength-deflection angle pairs, the best PGGAN-generated devices outperform those generated from the basic GAN.  This performance disparity can be further visualized by tracking device efficiencies over the course of network training for the PGGAN and basic GAN, which is summarized in Fig. \ref{fig4}(d).  For the basic GAN, the average efficiency of the generated devices initially increases but plateaus after approximately 20,000 iterations.  For the PGGAN, the average efficiency of the generated devices increases over the course of the full network training process, and upon training completion, PGGAN-generated devices have an average efficiency that is $30 \%$ higher than that generated form the basic GAN.  The superior performance of the PGGAN compared to the basic GAN is further enforced in Fig. \ref{fig4}(e), which shows efficiency histograms generated from the two methods.  We find that the PGGAN consistently generates devices with efficiency distributions that have peak and average values that are higher than those from the basic GAN. 

The ability for the PGGAN to generalize and accurately capture high resolution features from training data is due to the combination of progressive growth in both the training set and network architecture.  As a means of comparison, we have trained two alternative GANs, one with only progressive network architecture growth and the other with only progressive training set growth.  The results are summarized in Appendix B and indicate that progressive growth of either the network architecture or training set boosts the GAN performance beyond the basic GAN concept.  However, the performance from both of these GAN variants is still definitively worse than that of the full PGGAN.  

An estimation of the computational cost for the differing optimization methods featured in Figs. 4(a)-4(c) indicates that of the three options, the PGGAN concept requires the least computational overhead.  For the gradient-based topology-optimized devices in Fig. \ref{fig4}(a), 16.8M RCWA EM simulations are required. For the basic GAN devices in Fig. \ref{fig4}(b), EM simulations are required to produce the training set, evaluate the GAN-generated devices, and additionally refine the best GAN-generated devices \cite{Jiang19_acsnano}. In total, 4.5M RCWA EM simulations are required (see Appendix C). The PGGAN featured in Fig. \ref{fig4}(c) requires the production of the initial training set, evaluation of generated devices during network training to grow the training set, and the evaluation of final PGGAN-generated devices.  In total, 3.0M RCWA EM simulations are required. It is clear that PGGAN requires the fewest EM simulations, which is the primary factor determining computational cost, compared to the basic GAN and gradient-based topology optimization. Furthermore, most of the computational cost with the PGGAN is a one-time cost incurred during network training, such that a fully trained network only requires the evaluation of generated devices to produce high performance structures.

\section{Conclusion}
In summary, we present PGGAN as an effective and computationally efficient metasurface design methodology. Compared to the basic GAN, PGGAN benefits from the progressive growth of its network architecture, which enables more robust learning of topological features from the training set, and from the progressive growth of the training set, which enables further exploration of high-efficiency topological features in the design space. Our training results indicate that PGGAN can generate devices that outperform the original GAN by a wide margin and that compare well with those generated from gradient-based topology optimization, all while incurring  less computational cost.  Directions for  future work include identifying methods to reduce the computational cost of the training set and further optimizing the PGGAN architecture and a resolution transition scheme. We envision that PGGAN can be widely utilized in other fields that require the design of high-resolution, high-dimensional layouts with input parameters.  We also anticipate that our training set update strategy can be applied to other domains if a simulator exists that can evaluate the performance of generated layouts.

\appendix

\section{Protocol of PGGAN training set growth}
Every 10,000 training iterations, we consider wavelengths ranging from 500 nm to 1100 nm, in increments of 50 nm, and a deflection angle ranging from 35 degrees to 85 degrees, in increments of 5 degrees, and we generate 500 devices operating at each of these wavelength-deflection angle pairs. We then calculate the efficiencies of these devices using the RCWA EM solver. Devices that have efficiencies higher than the average efficiency of the training set, for a given wavelength-deflection angle pair, are added to the training set. 

We note that this sampling of deflection angles and wavelengths is relatively dense and that many of these wavelength-deflection angle pairs are not covered in the initial training set.  We add the five best PGGAN-generated devices for each of these wavelength-deflection angle pairs (not covered by the initial training set) to the training set during the first training set growth step.

Over the course of training, we also remove devices that have relatively low overall performance, which can impede the ability of the PGGAN to optimally train.  During each training set growth step, we identify the highest efficiency device in the training set for a given wavelength-deflection angle pair and we remove devices that have efficiencies lower than $85 \%$ of this highest efficiency value.  This method of training set refinement ensures that the performance of the training set, as quantified by the average device efficiency and size, improves over the course of network training.

\section{Individual training results of network growth and training set growth}

\begin{figure}[ht]
\centering
\includegraphics[width=0.8\linewidth]
{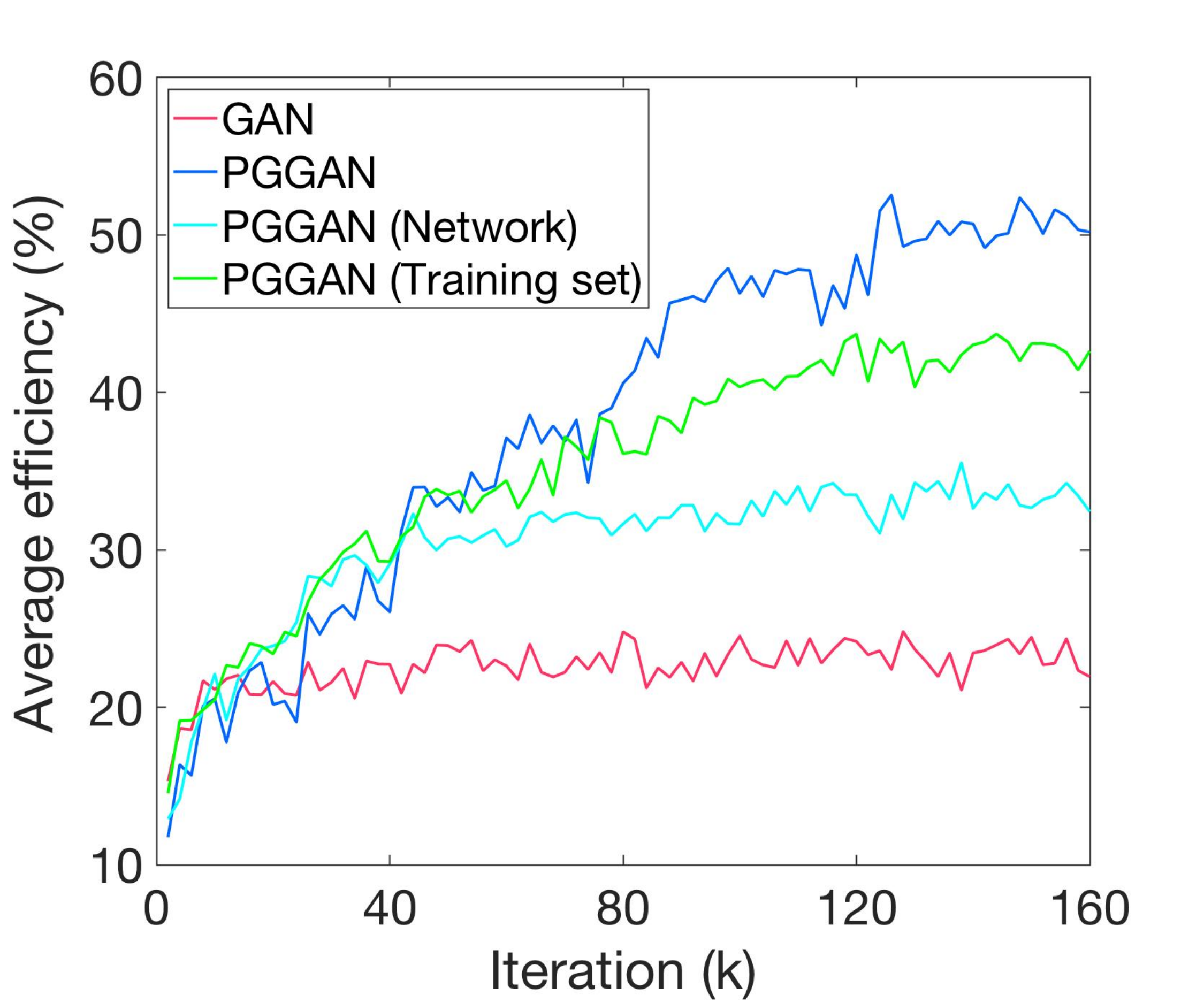}
\caption{
Average efficiencies of generated devices as a function of training iteration for teh basic GAN (red), PGGAN (blue), GAN with only network growth (green), and GAN with only training set growth (cyan). The addition of network growth boosts the average GAN-produce device efficiency by about $10 \%$, while the training set growth boosts the average GAN-produced device efficiency by about $20 \%$.
}
\label{fig5}
\end{figure}

The training results for the basic GAN, PGGAN, GAN with only training set growth, and GAN with only network growth are shown in Fig. \ref{fig5}. GAN performance benefits from network growth and training set growth, with training set growth contributing more than network growth. However, only the combination of both growth types can boost the performance of the GAN to levels comparable with gradient-base topology optimization.

\section{Computational cost calculation}
In this section, we further breakdown the computation cost of the optimization methods featured in Figure 5.  For steps requiring gradient-based topology optimization, six individual device simulations per iteration are used.  These include forward simulations of dilated, intermediate, and eroded patterns and their corresponding adjoint simulations.
To create the initial training set for the basic GAN and PGGAN, we  perform 350 iterations of adjoint-based topology optimization for 800 randomly generated devices and add the top $50\%$ devices (in terms of efficiency) to the training set. In total, the preparation of the sparse training set requires  $800\times350\times6=1,680,000$ simulations. 

1. For gradient-based topology optimization, 8000 devices are produced.  Each device needs 350 iterations of adjoint-based topology optimization, and it therefore takes $8,000\times350\times6 = 16,800,000$ simulations.

2. For the basic GAN, we generate 5000 devices for each wavelength and angle combination (13 wavelengths and 11 deflection angles), from which 50 best devices are chosen for 50 iteration of topology refinement. The computational cost for the basic GAN includes the preparation of the initial training set, the evaluation of the generated devices, and further generated device refinement.  In total, the number of required EM simulations is $1,680,000+13\times11\times5,000+13\times11\times50\times6\times50=4,540,000$ simulations.  Of these $4,540,000$ simulations, $13\times11\times5,000+13\times11\times50\times6\times50=2,860,000$ simulations are performed after the GAN model is trained.

3. For PGGAN, we perform 15 training set updates during network training.  For each training set update, we simulate the efficiencies of 500 devices at $13\times11$ combinations of wavelength and deflection angle. After training the PGGAN, we generate and evaluate 2000 devices for each wavelength and angle combination. The total computational cost includes the preparation of the initial training set, the training set updates, and the evaluation of the PGGAN-generated devices from a fully trained network.  The total number of simulations is: $1,680,000+13\times11\times15\times500+13\times11\times2,000=3,038,500$ simulations. Given a fully trained PGGAN, the number of simulations required to evaluate generated devices is $13\times11\times2000=286,000$ simulations.

\bibliography{refs}

\end{document}